\documentclass{iopart}
\usepackage{epsfig}
\begin{document}

\newcommand{\tbox}[1]{\mbox{\tiny #1}}
\newcommand{\half}{\mbox{\small $\frac{1}{2}$}}
\newcommand{\mbf}[1]{{\mathbf #1}}
\newcommand{\hide}[1]{}


\title{Non-perturbative response: chaos versus disorder}


\author{Doron Cohen$^{1}$ and Tsampikos Kottos$^{2}$}

\address{
\mbox{$^{1}$ Department of Physics, Ben-Gurion University, Beer-Sheva 84105, Israel} \\
\mbox{$^{2}$ Max-Planck-Institut f\"ur Str\"omungsforschung, Bunsenstra\ss e 10, D-37073 G\"ottingen, Germany}
 }




\begin{abstract}
Quantized chaotic systems are generically characterized by
two energy scales: the mean level spacing $\Delta$,
and the bandwidth $\Delta_b\propto\hbar$.
This implies that with respect to driving such systems
have an adiabatic, a perturbative, and a non-perturbative regimes.
A ``strong" non-linearity in the response, 
due to a quantal non-perturbative effect, 
is found for {\em disordered} systems that
are described by random matrix theory models.
Is there a similar effect for quantized {\em chaotic} systems?
Theoretical arguments cannot exclude the existence
of an analogous ``weak" version of the above mentioned 
non-linear response effect, but our numerics demonstrates 
an unexpected degree of semiclassical correspondence.
\end{abstract}


\section{Introduction}

\subsection{The two energy scales in Quantum chaos}

The name ``Quantum Mechanics" is associated with the idea
that the energy is quantized. For generic (chaotic) system
the mean level spacing is $\Delta\propto\hbar^d$,
where $d$ is the dimensionality of the system.
However, one should recognize that there is a second
energy scale $\Delta_b\propto\hbar$ which is introduced
by Quantum Mechanics. This $\hbar$ energy scale is
related to the chaos implied decay of the classical correlations.
It is known in the literature as the ``non-universal" energy
scale \cite{berry}, or as the ``bandwidth" \cite{mario}.
The dimensionless bandwidth is defined as $b=\Delta_b/\Delta$.
For reasonably small $\hbar$ one has $b\gg1$.

This observation, of having two energy scales, has motivated
the study of Wigner model \cite{wigner} within
the framework of random matrix theory (RMT). This model,
which is defined in terms of $\Delta$ and $\Delta_b$,
is totally artificial: it does not possess any classical limit.
Still note that it can be re-interpreted as a model for the motion
of a particle in a quasi one-dimensional {\em disordered} lattice~\cite{quasi}.

The main focus of Quantum chaos studies (so far) was on
issues of spectral statistics \cite{books}. In that context it turns
out that the sub-$\hbar$ statistical features of the
energy spectrum are ``universal", and obey the predictions
of random matrix theory. Non universal (system specific)
features are reflected only in the large scale properties
of the spectrum (analyzing energy intervals $>\Delta_b$).


\subsection{Regimes in the theory of driven systems}

In recent years we have made some progress in understanding
the theory of driven quantized chaotic systems \cite{crs,rsp,revw}.
Driven systems are described by Hamiltonian ${\cal H}(Q,P,x(t))$,
where~$x(t)$ is a time dependent parameter.
Such systems are of interest in mesoscopic physics (quantum dots),
as well as in nuclear, atomic and molecular physics.
The time dependent parameter $x(t)$ may have the significance
of external electric field or magnetic flux or gate voltage.
Linear driving $x(t)=Vt$ is characterized by one parameter ($V$),
while more generally a periodic driving $x(t)=A f(t)$ is characterized
by both amplitude ($A$) and frequency ($\Omega$).
Due to the time dependence of $x(t)$, the energy of the system
is not a constant of motion. Rather the system makes ``transitions"
between energy levels, and therefore absorbs energy.

The main object of our studies is the energy spreading
kernel $P_t(n|m)$. Regarded as a function of the level
index $n$, it gives the energy distribution after time $t$,
where $m$ is the initial level.
Having two quantal energy scales \mbox{($\Delta$, $\Delta_b$)}
implies the existence of different quantum-mechanical (QM)
$V$~{\em regimes} \cite{crs}, or more generally 
$(A,\Omega)$~{\em regimes} \cite{rsp}, in the theory of $P_t(n|m)$.
Most familiar is the QM adiabatic regime (very very small~$V$), 
whose existence is associated with having finite~$\Delta$.
Outside of the adiabatic regime we are used to the idea
that there is a perturbative regime, where the Fermi golden rule 
applies, leading to a Markovian picture of the dynamics,
with well defined transition rates between levels.
Less familiar \cite{crs,rsp} is the QM non-perturbative regime
($V$ is quantum mechanically large, but still classically small)
whose existence is associated with the energy scale~$\Delta_b$.
As implied by the terminology, in the QM non-perturbative regime
perturbation theory (to any order)  is not a valid tool for
the analysis of the energy spreading.
Consequently the Fermi golden rule picture of the dynamics 
does not apply there.


\subsection{Linear response theory}

Of special importance (see discussion below) is the 
theory for the variance $\delta E(t)^2=\sum_n P_t(n|m) (E_n-E_m)^2$ 
of the energy spreading. Having  $\delta E(t) \propto A$ 
means {\em linear response}. If $\delta E(t)/A$ depends on $A$ 
we call it ``non-linear response". In this paragraph we explain 
that linear response theory (LRT) is based on the ``LRT formula"  
\begin{eqnarray} \label{e1}
\delta E(t)^2 \ = \ A^2 \times
\int_{-\infty}^{\infty} \frac{d\omega}{2\pi}
\tilde{F}_t(\omega) \tilde{C}(\omega)
\end{eqnarray}
Two spectral functions are involved: One is the power
spectrum $\tilde{C}(\omega)$ of the fluctuations,
and the other $\tilde{F}_t(\omega)$  is the spectral
content of the driving. See Eq.(\ref{e4}) and Eq.(\ref{e5})
for exact definitions.
A special case of Eq.(\ref{e1}) is the sudden limit
($V=\infty$) for which $f(t)$ is a step function,
hence $F_t(\omega)=1$, and accordingly
\begin{eqnarray} \label{e2}
\delta E \ = \ \sqrt{C(0)} \times A
\hspace*{3cm} \mbox{\small [``sudden" case]}
\end{eqnarray}
Another special case is the response for
persistent (either linear or periodic) driving.
In such case the long time limit of
$F_t(\omega)$ is linear in time
[e.g. for linear driving ($f(t)=t$) we get
$F_t(\omega) = t \times 2\pi\delta(\omega)$].
This implies diffusive behavior:
\begin{eqnarray} \label{e3}
\delta E(t) = \sqrt{2 D_E t}
\hspace*{3cm} \mbox{\small [``Kubo" case]}
\end{eqnarray}
In the latter case the expression for $D_E\propto A^2$
is known as Kubo (or Kubo-Greenwood) formula,
leading to a fluctuation-dissipation relation \cite{revw}.

The LRT formula Eq.(\ref{e1}) has a simple classical
derivation\cite{crs}.
From now on it goes without saying that we
assume that the {\em classical} conditions
on $(A,\Omega)$ for the validity of Eq.(\ref{e1})
are satisfied (no $\hbar$ involved in such conditions).
The question is {\em what happens to the validity of
LRT once we ``quantize" the system}.
Can we trust Eq.(\ref{e1}) for any $(A,\Omega)$?
Or maybe we can trust it only in a restricted regime?
In previous publications\cite{crs,rsp,revw},
we were able to argue the following: \\
\begin{minipage}{\hsize}
\vspace*{0.2cm}
\begin{itemize}
\setlength{\itemsep}{0cm}
\item[(A)]
The LRT formula can be trusted
in the perturbative regime, with the exclusion
of the adiabatic regime.
\item[(B)]
In the sudden limit the LRT formula can
be trusted also in the non-perturbative regime.
\item[(C)]
In general the LRT formula cannot be
trusted in the non-perturbative regime.
\item[(D)]
The LRT formula can be trusted deep in the non-perturbative
regime, provided the system has a classical limit.
\end{itemize}
\vspace*{0.0cm}
\end{minipage}
For a system that does not have a classical limit
(Wigner model) we were able to demonstrate \cite{rsp}
that LRT fails in the non-perturbative regime. 
Namely, for Wigner model the response $\delta E(t)/A$ 
becomes $A$ dependent for large $A$, meaning that 
the response is non-linear. 
Hence the statement in item (C) above has been established.  
We had argued that the observed non-linear response     
is the result of a quantal non-perturbative effect.
{\em Do we have a similar type of non-linear response  
in case of quantized chaotic systems?} 
The statement in item (D) above seems to suggest that 
the observation of such non-linearity is not likely. 
Still, we argue below that this does not exclude 
the possibility of observing a ``weak" non-linearity.


\section{Perturbation theory and linear response}

The immediate (naive) tendency is to regard LRT as the outcome
of first order perturbation theory (FOPT).
In fact the regimes of validity of FOPT and of LRT
do not coincide. On the one hand we have the
adiabatic regime where FOPT is valid as a leading order
description, but not for response calculation
(see further details below).
On the other hand, the validity of Eq.(\ref{e1})
goes well beyond FOPT. This leads to the (correct) identification
\cite{crs,rsp} of what we call the ``perturbative regime".
The border of this regime [in $(A,\Omega)$ space]
is determined by the energy scale $\Delta_b$,
while $\Delta$ is not involved.
Outside of the perturbative regime we cannot
trust the LRT formula. However, as we further explain below,
the fact that Eq.(\ref{e1}) is not valid in the non-perturbative regime,
does not imply that it {\em fails} there.

We stress again that one should  distinguish 
between ``non-perturbative response" and ``non-linear response".
These are not  synonyms. As we explain in the next paragraph,
the adiabatic regime is ``perturbative" but ``non-linear",
while the semiclassical limit is ``non-perturbative" but ``linear".

In the {\em adiabatic regime}, FOPT implies zero probability
to make a transitions to other levels. Therefore, to the
extend that we can trust the adiabatic approximation,
all the probability remains concentrated in the initial level.
Thus, in the adiabatic regime, Eq.(\ref{e1}) is not a
valid formula: It is essential to use higher orders
of perturbation theory, and possibly non-perturbative
corrections (Landau-Zener \cite{wilk}), in order to
calculate the response. Still, FOPT provides a meaningful 
leading order description of the dynamics (i.e. having no transitions), 
and therefore we do not regard the adiabatic
non-linear regime as ``non-perturbative".

In the {\em non-perturbative regime} the evolution
of $P_t(n|m)$ cannot be extracted from perturbation theory:
not in leading order, neither in any order.
Still it does not necessarily imply a non-linear response.
On the contrary: the semiclassical limit is contained in
the (deep) non-perturbative regime\cite{rsp}.
There, the LRT formula Eq.(\ref{e1}) is in fact valid. But its
validity is {\em not} a consequence of perturbation theory, 
but rather the consequence of {\em quantal-classical correspondence}.


\section{The quest for non perturbative response}

As stated above, an effect of non-linear response 
due to the {\em quantum mechanical} non-perturbative nature 
of the dynamics, has been demonstrated so far only 
for Wigner model \cite{rsp}.
There, its existence is related to the {\em disordered} RMT nature
of the model (see discussion below).
Semiclassical correspondence considerations
seem to exclude the manifestation of this disorder-related
non-linearity in case of quantized {\em chaotic} systems.
In this Letter we explain that this does not exclude
the possibility of having a ``weak" version of this effect.
We also report the results of an intense numerical effort
aimed in finding a ``weak" non-linearity in the case of a simple
low-dimensional quantized chaotic systems. To our surprise,
an unexpected degree of semiclassical correspondence is observed.

It is appropriate here to clarify the notions of ``weak"
and ``strong" effects. In the literature regarding the
dynamics in disordered lattices one distinguishes between ``weak"
and ``strong" localization effects. The former term implies
that while the leading behavior is classical (diffusion),
there are ``on top" quantum mechanical corrections
(enhanced return probability). In contrast to that the
term ``strong" implies that the classical description fails
even as a leading order description. In the literature
regarding quantum chaos we have the effect of ``scarring",
which should be regarded as ``weak" effect.
"Strong" quantum mechanical effects (e.g. dynamical localization
in 1D kicked systems \cite{qkr}) are non-generic:  The leading order
behavior of generic quantized chaotic systems is typically classical.
In the present context of driven systems, we use the
terms ``weak" and ``strong" in the same sense: 
The adjective ``weak" is associated with the (conjectured) 
non-linear response of quantized driven chaotic systems,
while the adjective ``strong" is associated
with the (established) non-linear response
in the corresponding RMT (Wigner) model.


\section{The numerical findings}

How do we detect non-linear response?
The most straightforward way is to fix the pulse shape
$f(t)$ and to plot $\delta E/A$ versus $A$.
A deviation from constant value means
"non-linear" response. The simulations below
are done for a quantized chaotic system.
Due to obvious numerical limitations
we will consider the response to one-pulse driving,
rather than to persistent (periodic) driving.
{\em The central question is whether the observed non-linear
effect is of semiclassical origin, or of novel
quantum mechanical origin}. We deal with this issue below.

In our numerical simulations (Fig.1) we have considered
a particle in a two dimensional ($d=2$) anharmonic well.
This model (with deformation parameter $x=\mbox{const}$)
is defined in \cite{lds,wpk}.
In the energy region of interest ($E\sim3$), the classical
motion inside the two dimensional well (2DW) is chaotic,
with characteristic correlation time $\tau_{cl}\sim 1$.
For the following presentation it is enough to say that
the quantum mechanical Hamiltonian is represented
by a matrix ${\cal H}=\mbf{E}+x(t)\mbf{B}$,
where $\mbf{E}$ is a diagonal matrix with mean
level spacing $\Delta\approx 4.3\times\hbar^{d}$,
and $\mbf{B}$ is a banded matrix.
The bandwidth (in energy units)
is $\Delta_b=2\pi\hbar/\tau_{cl}$.
The bandprofile (see Fig.2 of \cite{lds}) is described
by a spectral function which is defined as follows:
\begin{eqnarray} \label{e4}
\tilde{C}(\omega) \ = \
\sum_{n(\ne m)}
\left|\mbf{B}_{nm}\right|^2
\ 2\pi\delta\left(\omega-\frac{E_n{-}E_m}{\hbar}\right)
\end{eqnarray}
with implicit average over the reference state $m$.
The bandprofile, as defined above, can be determined
from the the classical dynamics. This means
that $\tilde{C}(\omega)\approx\tilde{C}^{cl}(\omega)$
where $\tilde{C}^{cl}(\omega)$ is the Fourier transform
of a classical correlation function $C^{cl}(\tau)$.
The $\hbar$ dependence of $\tilde{C}(\omega)$ is relatively weak.

The driving pulse in our numerical simulations
has a rectangular shape. This means $f(0)=f(T)=0$
and $f(0<t<T)=A$, where $T=0.375$.
The spectral content of the driving is defined as:
\begin{eqnarray} \label{e5}
\tilde{F}_t(\omega) \ = \
\left| \int_0^t \dot{f}(t') \mbox{e}^{i\omega t'} dt' \right|^2
\end{eqnarray}
The spectral content of the driving after a rectangular pulse
is  $F_t(\omega) = |1-\mbox{e}^{i\omega T}|^2$.
We have also made simulations (not presented) with a driving
scheme that involves a positive pulse $+A$ followed by
a negative pulse $-A$, with the intention of considering
eventually a persistent (multi cycle) periodic driving.
However, we have realized that all the relevant
physics is observed already in the single pulse case.
Note that the regime diagram for either
linear or (as in the following simulations)
rectangular driving pulse, is greatly simplified,
because the driving is characterized by only one parameter
($V$ in the former case, $A$ in the latter case).

Let us look carefully at the results of the 2DW simulations (Fig.1).
For small $A$ we see as expected ``linear response"
meaning $\delta E /A = \mbox{const}$, as implied by Eq.(\ref{e1}).
Note that the ``constant" has a weak $\hbar$ dependence (a $10\%$ effect).
This is due to the above mentioned weak dependence of $\tilde{C}(\omega)$
on $\hbar$. So this quantum-mechanical effect is quite
trivial, and has a simple explanation within LRT.   
Now let us look what happens for large $A$.
We clearly see a $2\%$~deviation from linear response. 
In what follows we discuss the reason for this non-linear effect.

For sake of comparison we also perform simulations with an
effective RMT model that corresponds to the 2DW model Hamiltonian.
The effective RMT model is obtained by randomizing the signs
of the off-diagonal elements of the $\mbf{B}$ matrix.
The effective RMT Hamiltonian has the {\em same} bandprofile
$\tilde{C}(\omega)$ as the original (2DW) Hamiltonian.
Therefore, as far as LRT Eq.(\ref{e1}) is concerned, the response should
be the {\em same}. Still we see that at the same $A$ regime,
as in the case of the 2DW simulations, we have deviation
from linear response. However, this non-linear deviation 
is much much stronger.

Looking at the curves of Fig.1, it is very tempting to regard the observed
non-linear $2\%$ effect in the 2DW simulations as a ``weak" version
of the ``strong" effect which is observed in the corresponding RMT simulations.
However, the careful analysis below indicates that apparently
this is not the case.


\section{Discussion and analaysis}

In analyzing the validity of the LRT formula,
it is instructive to consider first the
sudden limit Eq.(\ref{e1}). This limit has been
studied in \cite{lds}. The spreading profile
$P(n|m)$, after the sudden change in $x$,
depends on the amplitude $A$ of the
perturbation. [We omit the time index $t$,
which is of no relevance in this limit].
The perturbative regime is $A<A_{\tbox{prt}}$,
where $A_{\tbox{prt}}=2\pi\hbar/(\tau_{cl}\sqrt{C(0)})$.
For the 2DW simulations $A_{\tbox{prt}} = 5.3\times\hbar$.
In the perturbative regime $P(n|m)$
has a core-tail structure
(the generalization of Wigner Lorentzian),
and the variance $\delta E^2$ is determined
by the first order tail component of the energy distribution.
For $A>A_{\tbox{prt}}$ the spreading profile
$P(n|m)$  becomes non-perturbative.
This means that the perturbative tail (if survives)
is no longer the predominant feature. Thus the variance
is determined by the non-perturbative component
(the ``core") of the energy distribution.
The remarkable fact is that
the crossover from the perturbative $A$ regime
to the non-perturbative $A$ regime is {\em not}
reflected in the variance (see Fig.5 of \cite{lds}).
The agreement with Eq.(\ref{e2}) is perfect.
Taking into account the ``dramatic" differences
in the appearance of $P(n|m)$, this looks
quite surprising. In fact (see Sec.12 of \cite{lds})
there is a simple proof \cite{felix} that
Eq.(\ref{e2}) remains exact beyond any order
of perturbation theory, which means that it
is exact even in the non-perturbative regime
where perturbation theory is not applicable.

We turn back to our simulations,
where we have a rectangular pulse
(rather than step function). Here
the sudden limit does not apply,
and the dynamics within the time
interval $0<t<T$ should be taken
into account.
If we take an eigenstate of $\mbf{E}$ and propagate
it using $\mbf{E}+A\mbf{B}$, then we get in the classical case
{\em ballistic} spreading followed by saturation.
["Eigenstate" in the classical case means microcanonical
distribution]. This is true for any $A$.
Quantum mechanically we observe in the 2DW model
simulations a similar ballistic behavior \cite{wpk},
whereas in the corresponding RMT model there is
an intermediate stage of {\em diffusion} \cite{wpk}.
This diffusion is of non-perturbative nature,
and it is related to the ``disorder" which is artificially
introduced via the sign-randomization procedure.
The strong non-linear response effect \cite{rsp}
is a consequence of this diffusion.

Coming back to the 2DW model, we realize that there is
no ``disorder" build into the model, and therefore
no diffusion. Still, looking at Fig.1,
it is tempting to interpret the observed
$2\%$~non-linear deviation as a ``weak" version of the
strong non-linear effect.
Moreover, regarded as such, it vanishes, as expected,
in the deep non-perturbative regime,
which had been argued on the basis of
semiclassical correspondence considerations\cite{rsp}.

In order to properly determine whether the dips
in Fig.1 are the result of the QM non-perturbative 
nature of the dynamics, we have rescaled the vertical axis,
and plotted  the response once (Fig.2a) versus $A$,
and once (Fig.2b) versus $A/\hbar$.
On the basis of  the scaling we see that
the strong non-linear response in the RMT
case is indeed the result of the quantal ($\hbar$-dependent)
non-perturbative effect. In contrary to that
the $\hbar$-independent scaling in the 2DW case,
indicates that the non-linear deviation there is of ``semiclassical"
rather than ``quantal non-perturbative" nature.


\section{Conclusion}

Theoretical arguments cannot exclude
the existence of a ``weak" non-linearity  
in the response of a driven quantized chaotic system, 
which is due to a quantum mechanical non-perturbative effect.
But our careful numerics, regarding a simple
low-dimensional system, demonstrates an unexpected
degree of semiclassical correspondence.
Our findings should be regarded as the outcome
of an ongoing quest, which has not ended,
that is aimed in finding novel quantum mechanical
deviations from linear response theory.

\ \\

This work was supported by a grant from the GIF,
the German-Israeli Foundation for Scientific Research and Development,
and by the Israel Science Foundation (grant No.11/02).

\newpage


\newpage

\centerline{\epsfig{figure=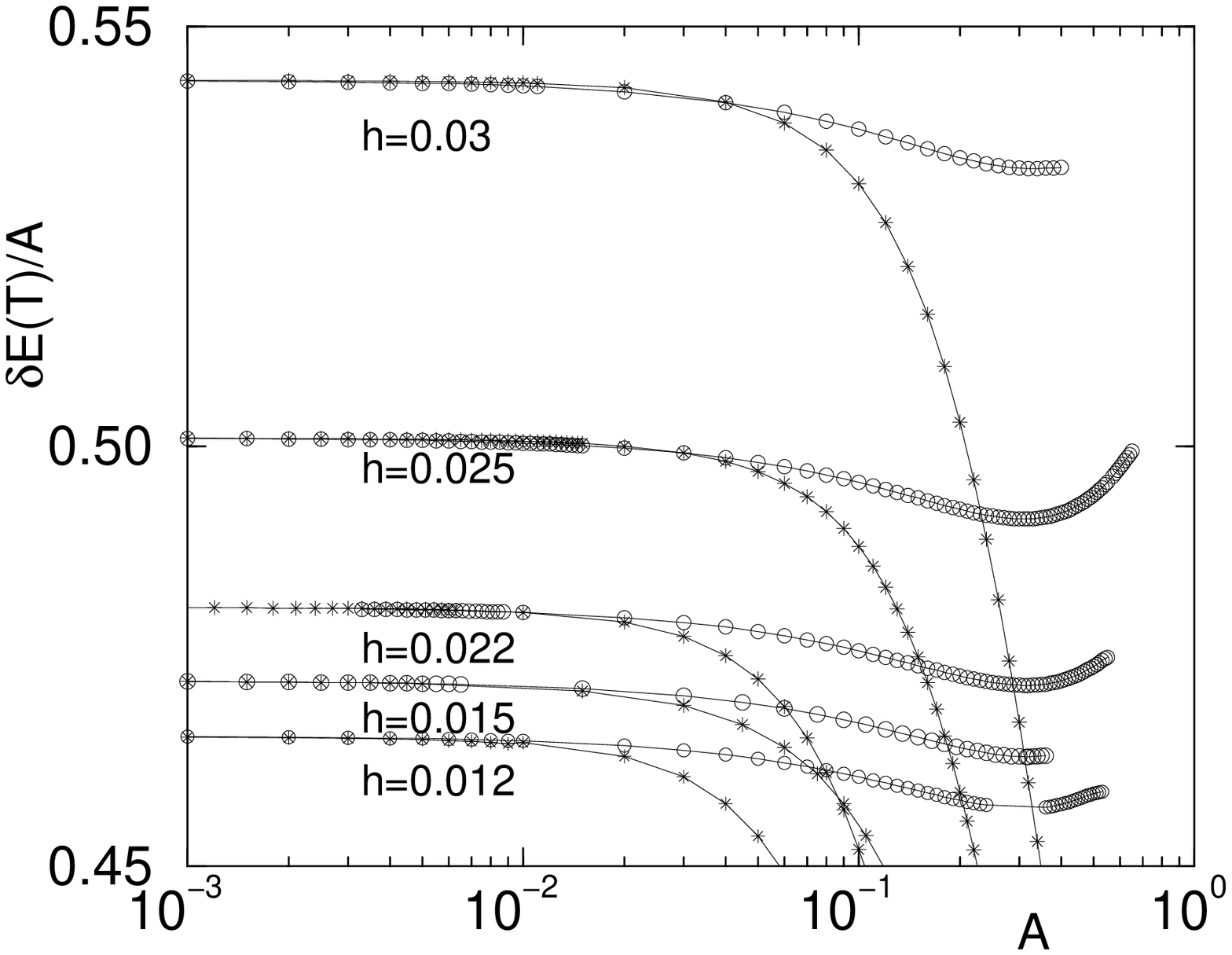, width=0.7\hsize}}
{\footnotesize FIG.1:
The response $\delta E/A$ as a result of a rectangular
pulse ($T=0.375$). Deviation from $\delta E/A=\mbox{const}$
implies non-linear response. All the data are averaged
over a number of different initial conditions. The simulations
are done with the 2DW Hamiltonian (circles), and also with the
associated RMT model (stars). See text for explanations.}

\ \\

\centerline{\epsfig{figure=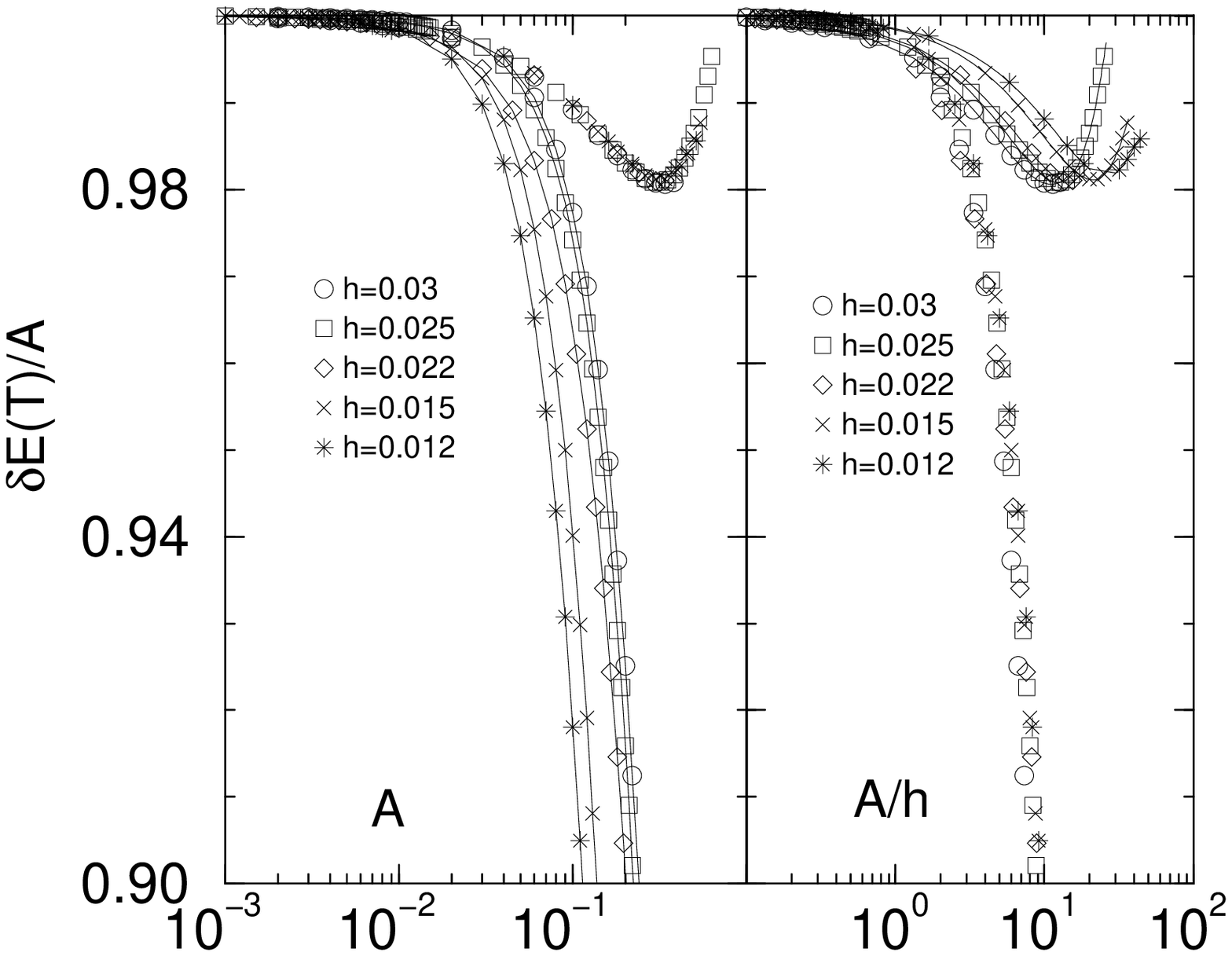, width=0.8\hsize}}
{\footnotesize FIG.2: Scaled versions of Fig.1. The vertical
scaling is aimed in removing the weak $\hbar$ dependence
of the bandprofile. In (b) the horizontal scaling is aimed in
checking  whether the deviation from linear response
is in fact a quantal non-perturbative effect.} 

\end{document}